\documentclass[twocolumn,twoside,slac_two]{revtex4}
\usepackage{graphicx}
\usepackage{fancyhdr}
\input babarsym

\def\journal#1#2#3#4{#1~{\bf #2}, #3 (#4)}
\def\PL#1#2#3{\journal{Phys.\ Lett.}{#1}{#2}{#3}}
\def\NP#1#2#3{\journal{Nucl.\ Phys.}{#1}{#2}{#3}}
\def\PR#1#2#3{\journal{Phys.\ Rev.}{#1}{#2}{#3}}
\def\PRL#1#2#3{\journal{Phys.\ Rev. Lett.}{#1}{#2}{#3}}

\newcommand{\etal}{{\em et al.}}

\begin{document}

\title{CKM matrix fits including constriants on New Physics}

%

\author{H. Lacker}
\affiliation{Institut f\"ur Kern- und Teilchenphysik, Technische Universit\"at Dresden, 
01062 Dresden, Germany}

\begin{abstract}
I review the status of global fits to the CKM matrix within the framework 
of the Standard Model and also allowing for New Physics contributions in 
$B-\bar{B}$ mixing. The driving force is coming from the large data sets 
collected by the $B$-factory experiments \babar\ and Belle. 
Additional important inputs to the $B_{s}$ sector are provided by the Tevatron 
experiments CDF and D0. In particular, when constraining New Physics in 
$B-\bar{B}$ mixing in a model-independent analysis a nice interplay between 
the $B$-factories and the Tevatron experiments is observed.
\end{abstract}

\maketitle

\thispagestyle{fancy}


\section{Introduction}  
Within the Standard Model (SM) quark flavor-mixing is described by the 
$3 \times 3$ unitary Cabibbo-Kobayashi-Maskawa (CKM) matrix~\cite{CKM}. 
The size of \CP\ violation carried by the CKM matrix is proportional to 
a single parameter, the Jarlskog invariant 
$J={\Im}\left[V_{ij}V_{kl}V_{il}^*V_{kj}^*\right]$~\cite{Jarlskog}, where
$|J|/2$ quantifies the area of the unitarity triangle (UT), defined by
$V_{ud} {V_{ub}}^{*} + V_{cd} {V_{cb}}^{*} + V_{td} {V_{tb}}^{*} = 0$. 
 
The CKM matrix is parametrized by four independent real parameters. 
Inspired by the experimentally observed hierarchy of the CKM matrix, 
``Wolfenstein-type'' parametrizations have been proposed in the 
literature (see e.g. Refs.~\cite{Wolfenstein,BurasLautenbacherOstermaier,CKMfitter}). 
Based on the improved Wolfenstein 
parametrization~\cite{BurasLautenbacherOstermaier} the following set 
of CKM matrix parameters has been proposed~\cite{CKMfitter} and is 
advertized by the PDG group~\cite{PDG2006}:
$\lambda=\frac{|V_{us}|}{\sqrt{{|V_{ud}|}^2+{|V_{us}|}^2}}$, 
$A \lambda^2=\frac{|V_{cb}|}{\sqrt{{|V_{ud}|}^2+{|V_{us}|}^2}}$, and 
$\rhobar+i \cdot \etabar = -V_{ud}V_{ub}^{*}/V_{cd}V_{cb}^{*}$. 
This parametrization has several convenient properties: it is exact, 
it is unitary to all orders of $\lambda$, and it is phase-convention 
independent where (\rhobar,\etabar) represent the coordinates of the
apex of the rescaled UT 
($(V_{ud}{V_{ub}}^{*}+V_{cd}{V_{cb}}^{*}+V_{td}{V_{tb}}^{*})/V_{cd}{V_{cb}}^{*}=0$).

In many theoretical extensions of the SM sizeable New Physics (NP) effects 
are expected to contribute to the $B_{q}-\bar{B}_{q}$ mixing amplitude 
where $q=d,s$ (see e.g.~\cite{fleischerisidorijamin}). 
Since $\Gamma_{12}$ is dominated by long-distance physics one usually 
assumes that only $M_{12}$, which is dominated by intermediate top-quark 
contributions, is affected by NP. In this case and by assuming a 
$3 \times 3$ unitary CKM matrix one can parameterize NP in mixing by two 
new parameters where two different parametrizations are typically in use: 
$r_{q}^2 e^{2 i \theta_{q}} = \frac{<{\bar{B}}^{0}_{q}|M_{12}^{SM+NP}|B^{0}_{q}>}{<{\bar{B}}^{0}_{q}|M_{12}^{SM}|B^{0}_{q}>}$~\cite{rtheta}, 
respectively, 
$h_{q} e^{2 i \sigma_{q}} = \frac{<{\bar{B}}^{0}_{q}|M_{12}^{NP}|B^{0}_{q}>}{<{\bar{B}}^{0}_{q}|M_{12}^{SM}|B^{0}_{q}>}$~\cite{Goto,Agashe}.
The SM values are $r_{q}^2 =1$ and $2 \theta_{q} = 0$, respectively,
$h_{q} = 0$ and $2 \sigma_{q} = 0$.
 
The case of $K-\bar{K}$ mixing is more involved. The observable $\epsilon_K$ 
obtains sizeable contributions from the charm quark box diagram while the 
short distance contribution to $\Delta m_{K}$ is even dominated by intermediate 
charm quarks with a subdominant top quark contribution and a large, but hardly 
calculable long distance contribution from intermediate $u$ quarks. 
If one assumes that NP does only modify the top quark amplitude the observable 
$\Delta m_{K}$ provides only a weak constraint on the parameter $r_{K}^{2}$. 
In Ref.~\cite{Agashe} the leading top-quark contribution in $\epsilon_K$ has
been modified by a term containing the parameters $h_{K}$ and $2 \sigma_{K}$.
The UTfit collaboration simply parametrized any deviation from the SM in 
$\epsilon_K$ by $\epsilon_{K}^{exp}=C_{\epsilon_K} \cdot \epsilon_{K}^{SM}$ 
where 
$C_{\epsilon_K}=\frac{\Im<{\bar{K}}^{0}|H_{12}^{SM+NP}|K^{0}>}{\Im<{\bar{K}}^{0}|H_{12}^{SM}|K^{0}>}$~\cite{UTfit1}. 

Often, the additional assumption is made that NP does not contribute to 
tree-mediated decays. However, on the non-perturbative level tree and 
penguin amplitudes can not be well separated. For this reason, it has 
been proposed to be more precise that decays proceeding through a Four 
Flavor Change obtain only SM contributions ($SM4FFC$)~\cite{CKMfitter,Goto}. 
According to these assumptions the following inputs used in the fit allowing 
for NP in mixing are considered to be free from NP contributions in their
extractions from data: e.g. $|V_{ud}|$, $|V_{us}|$, $|V_{ub}|$, $|V_{cb}|$ 
and $\gamma$. Observables that are affected by NP in mixing are shown in 
Table~\ref{tab:observables}.
\begin{table}[]
\renewcommand{\arraystretch}{1.3}
\centering
\begin{tabular}{|c|c|}\hline
Observable           & Prediction in the presence of NP in mixing \\
\hline
$S(\psi K_{S})$      & $\sin{(2\beta + 2\theta_{d})}$             \\
$B \to \psi K^{0*}$  & $\cos{(2\beta + 2\theta_{d})}$             \\
$\alpha$             & $\pi - \gamma - \beta -\theta_{d}$         \\
$\Delta m_{q}$       & $\Delta m_{q}^{SM} \cdot r_{q}^{2}$        \\
$\Delta \Gamma_{s}$  & $-\Delta m_{q}^{SM} \left[\Re{\frac{\Gamma_{12}^{SM}}{M_{12}^{SM}}} \cos{2\theta_{q}}+\Im{\frac{\Gamma_{12}^{SM}}{M_{12}^{SM}}} \sin{2\theta_{q}}\right]$ \\
$A_{SL}^{q}$         & $-\Re{\frac{\Gamma_{12}^{SM}}{M_{12}^{SM}}} \frac{\sin{2\theta_{q}}}{r_{q}^{2}}                 
                        +\Im{\frac{\Gamma_{12}^{SM}}{M_{12}^{SM}}} \frac{\cos{2\theta_{q}}}{r_{q}^{2}}$ \\
\hline
\end{tabular}
\caption[Observables]
{Theoretical prediction of observables in the $B$-meson system in the 
presence of NP in mixing.}
\label{tab:observables}
\end{table}
It should be noted that 
$\Im{\frac{\Gamma_{12}^{SM}}{M_{12}^{SM}}}$ is much smaller than
$\Re{\frac{\Gamma_{12}^{SM}}{M_{12}^{SM}}}$, hence, terms containing 
$\Im{\frac{\Gamma_{12}^{SM}}{M_{12}^{SM}}}$  are often neglected in
the expressions given in Table~\ref{tab:observables}.

The first analysis making use of the early $B$-factory data ($|V_{ub}|$, $|V_{cb}|$, 
$\Delta m_{d}$, $\sin{(2\beta)}$) with a special focus on the role of $A_{SL}^{d}$ 
has been described in Ref.~\cite{Laplace}. The first complete model-independent 
analysis on data in the $B_{d}$ sector using all relevant observables ($|V_{ub}|$, 
$|V_{cb}|$, $\Delta m_{d}$, $\sin{(2\beta)}$, $\cos{(2\beta)}$, $\alpha$, $\gamma$, 
$A_{SL}^{d}$) is discussed in Ref.~\cite{CKMfitter} and could exclude a real CKM 
matrix even in the presence of NP in $B_{d}$ mixing.
A combined analysis for the Kaon and $B_{d}$ sector with prospects in the $B_s$ sector 
has been performed subsequently in Ref.~\cite{Agashe} examining a Next-To-Minimal 
Flavour Violation (NMFV) scenario, and for the Kaon and $B_{d}$ sector by the UTfit 
collaboration~\cite{UTfit1} with a focus on Minimal Flavour Violation (MVF). 
The advent of the first observation of $B_{s}$ oscillations by CDF~\cite{DmsCDF1} 
triggered several analyses 
(see e.g.~\cite{Blanke,BallFleischer,LigetiPapucciPerez,GrossmanNirRaz,UTfit2}) where 
not only $\Delta m_{s}$ but also the role of $A_{SL}$ in the $B_{d}$ and  $B_{s}$ 
sector as well as $\Delta \Gamma_{s}$ was discussed in several publications (see e.g. 
Refs.~\cite{LigetiPapucciPerez,GrossmanNirRaz,UTfit2}).

\section{Inputs}
\label{sec:inputs}
In this section, the status of experimental measurements and theoretical input 
parameters as used in state-of-the-art CKM fits is summarized. The numerical 
values are given in Table~\ref{tab:inputs}.

\subsection{Inputs free from New Physics in mixing}  
Currently, the best determination of $|V_{ud}|$ comes from superallowed
$\beta$-decays where the uncertainty is dominated by the theoretical
error, see e.g. Ref~\cite{CKM05Vud}. The matrix element $|V_{us}|$ is 
determined from $K_{e3}$ decays, from the ratio of rates between 
$K \to \mu \nu$ and  $\pi \to \mu \nu$, from $\tau$ decays, and from 
semileptonic hyperon decays. 
In his Moriond 2007 review talk, M.~Jamin quotes an average value of 
$|V_{us}|=0.2240 \pm 0.0011$~\cite{jamin}. It should be noted that this 
average is dominated by the $K_{e3}$ number where the quoted average has 
a significantly smaller theoretical uncertainty than quoted by others. 
This number is dominated by a recent and preliminary Lattice QCD (LQCD) 
calculation for the $K \to \pi$ form factor $f_{+}$ which has not evaluated 
all systematic uncertainties yet~\cite{VusUKQCDRBC}.
When using this $|V_{us}|$ input the uncertainty on $\lambda$ obtains
similar contributions from $|V_{ud}|$ and $|V_{us}|$ resulting in a 
$2\sigma$ deviation from the unitarity condition in the first family. 
When discarding this specific form factor calculation the weight of 
the $K_{e3}$ number would get smaller in the $|V_{us}|$ average but 
the size of the unitarity violation would remain more or less the same 
since the $|V_{us}|$ values from the other methods give smaller results.

The matrix element $|V_{cb}|$ is obtained from semileptonic decays
$B \to X_{c} \ell \nu$ where $X_c$ is either a $D^{*}$ meson (exclusive 
method) or a sum over all hadronic final states containing charm 
(inclusive method). The most precise value is currently provided by the 
inclusive method where the theoretical uncertainties have been pushed 
below the $2~\%$ level by determining non-perturbative parameters
from moment measurements in $B \to X_{c}\ell\nu$ and $B \to X_{s}\gamma$. 
The inclusive $|V_{cb}|$ value used in this analysis is taken from 
Ref.~\cite{BuchmuellerFlaecher}. 
The theoretical uncertainty on the exclusive $|V_{cb}|$ determination in 
the calculation of the form factor value $F$ at zero recoil is currently 
not competitive with the inclusive method. The central value is smaller
than but compatible with the inclusive result given its large theoretical 
uncertainty.

A delicate input is $|V_{ub}|$ for several reasons: It plays a crucial 
role in testing a NP phase in $B_{d}-\bar{B}_{d}$ mixing which can be 
detected by comparing the measured $\sin{2\beta}$ value with the 
predicted value without using the experimentally measured $\sin{2\beta}$ 
value. The predicted value is in particular sensitive to $|V_{ub}|$ since 
the $\sin{2\beta}$ constraint (in the SM) is tangent to the $|V_{ub}/V_{cb}|$ 
circle in the $\rhobar$-$\etabar$ plane.
However, the two methods to extract $|V_{ub}|$, the inclusive and the 
exclusive (using $B \to \pi \ell \nu$) are suffering from significant 
theoretical uncertainties and, in addition, do not perfectly agree with 
each other.
The exclusive numbers prefer values below $4.0 \times 10^{-3}$\footnote{
After the conference the HPQCD collaboration submitted an erratum for their 
former published Lattice QCD (LQCD) result for $f_{+}$. As a consequence, 
all LQCD values and the one from Light Cone Sum Rules (LCSR) prefer now 
values close to or even below $3.5 \times 10^{-3}$ which leads to an even 
more pronounced discrepancy.}. 
For the fit input the exclusive numbers quoted by the Heavy Flavour Averaging 
group (HFAG)~\cite{HFAG06} I average them in a conservative way by keeping the 
smallest theoretical uncertainty as a common theoretical error:
$|V_{ub,excl}|=(3.60\pm0.10\pm0.50) \times 10^{-3}$.
The average of inclusive results quoted by HFAG using e.g. the Shape 
Function (SF) scheme~\cite{BLNP} yields $(4.52\pm0.19\pm0.27) \times 10^{-3}$ 
where the first uncertainty contains the statistical and experimental 
systematic uncertainty as well as the modelling errors for $b \to u \ell \nu$ 
and $b \to c \ell \nu$ transitions.
Since several theoretical uncertainties are somehow guestimated (the HQE 
error on the $m_{b}$ mass determined in moment fits, the matching scale in 
the BLNP calculation, subleading shape functions and weak annihilation) 
I add those linearily and use $(4.52\pm0.23\pm0.44) \times 10^{-3}$ as an input 
where the first error is considered as a statistical one and the second as 
a theoretical one which will be scanned in the CKM fit
\footnote{
According to M. Neubert the uncertainty due to the SF parameters determining
the first and the second moment of the SF are underestimated in the HFAG
2006 average since the BLNP generator used to calculate partial rates is 
calculated on one-loop level. Hence the b-quark mass in the SF scheme has 
already an intrinsic uncertainty of $O(60-70)$ MeV which is significantly 
larger than the one obtained from moment fits~\cite{BuchmuellerFlaecher}. 
As a consequence, in future averages the uncertainty on $|V_{ub,incl}|$ is 
expected to become larger.}.
The inclusive and exclusive number are averaged again by keeping the smallest 
theoretical uncertainty as the common theoretical uncertainty 
(see Table~\ref{tab:inputs}). 

The UTfit group interprets all HFAG uncertainties in a statistical way and 
use the following inputs: $|V_{ub,incl}^{UTfit}|=(4.49\pm0.33) \times 10^{-3}$ 
and $|V_{ub,excl}^{UTfit}|=(3.50\pm0.40) \times 10^{-3}$.
If all uncertainties were interpreted as coming from Gaussian distributions
one obtained as an average $|V_{ub}|=(4.09\pm0.25) \times 10^{-3}$. If one
followed the PDG error rescaling recipe one would obtain 
$|V_{ub}|=(4.09\pm0.49) \times 10^{-3}$). 
As a result of the error treatment the discrepancy between the $|V_{ub}|$ 
input and its prediction is much more pronounced in the UTfit analysis than
in the analysis presented here.

The input for the UT angle $\gamma$ is taken from a combined full frequentist
analysis of the CKMfitter group using $CP$ violating asymmetries in charged $B$ 
decays to neutral $D^{(*)}$ mesons plus charged $K^{(*)}$ mesons as discussed 
in detail at this conference~\cite{Vincent}. On the $68~\%$ confidence level 
(CL) the result is $(77 \pm 31)^{\circ}$. This constraint differs significantly 
from the result obtained from the same data set by the UTfit group using a 
Bayesian approach ($(82 \pm 20)^{\circ}$ at $95~\%$ CL).
\begin{table}[]
\renewcommand{\arraystretch}{1.3}
\centering
\begin{tabular}{|c|c|}\hline
Observable                        & Value and Uncertainties                                             \\
\hline
$|V_{ud}|$                        & $0.97377 \pm 0.00027$                                               \\
$|V_{us}|$                        & $0.2240  \pm 0.0011$~\cite{jamin}                                   \\
$|V_{cb}|$                        & $0.0416  \pm 0.0007$, see text                                      \\
$|V_{ub}|$                        & $(4.09 \pm 0.09\pm 0.44) \times 10^{-3}$, see text                         \\
$\gamma$                          & $B^{\pm} \to D^{(*)} K^{(*)\pm}$, see text                          \\
$\alpha$                          & $B \to \pi\pi, \rho\rho, rho\pi$, see text                          \\
$\sin 2\beta$                     & $0.678   \pm 0.025$~\cite{HFAG06}                                   \\
$\cos 2\beta$                     & see text                                                            \\
$\Delta m_{d}$                    & $(0.507  \pm 0.005)~{\rm ps^{-1}}$~\cite{PDG07}                     \\
$\Delta m_{s}$                    & $(17.77  \pm 0.12)~{\rm ps^{-1}}$~\cite{DmsCDF2}                    \\
$A_{SL}^{d}$                      & $-0.0043 \pm 0.0046$, see text                                      \\
$A_{SL}$                          & $-0.0028 \pm0.0013 \pm 0.0008$~\cite{ASLD0}                         \\
$A_{SL}^{s}$                      & $0.0245  \pm 0.0196$\cite{ASLsD0}                                   \\
$\Delta \Gamma^{CP'}_{s}$         & $(0.12   \pm 0.08)~{\rm ps^{-1}}$                                   \\
$\epsilon_{K}$                    & $(2.232  \pm 0.007) \times 10^{-3}$~\cite{PDG2006}                         \\
\hline
$f_{B_{s}}$                       & $(268    \pm 17   \pm 20)~{\rm MeV}$~\cite{Tantalo}                 \\
$B_{s}$                           & $1.29    \pm 0.05 \pm 0.08$~\cite{Tantalo}                          \\
$f_{B_{s}}/f_{B_{d}}$             & $1.20    \pm 0.02 \pm0.05$~\cite{Tantalo}                           \\
$B_{s}/B_{d}$                     & $1.00    \pm 0.02$~\cite{Becirevic}                                 \\
$\eta_{B}$                        & $0.551   \pm 0.007$~\cite{BuchallaBurasLautenbacher}                \\
$m_{t}(m_{t})$                    & $(163.8  \pm 2.0)~{\rm GeV}$~\cite{Nierste3}                        \\
${\hat{B}}_{K}$                   & $(0.78   \pm 0.02 \pm 0.09)$                                        \\
$\eta_{tt}$                       & $0.5765  \pm 0.0065$~\cite{Nierste1,Nierste2}                       \\
$\eta_{ct}$                       & $0.47    \pm 0.04$~\cite{Nierste1,Nierste2}                         \\
$\eta_{cc}$                       & see text~\cite{Nierste1,Nierste2}                                            \\
$m_{c}(m_{c})$                    & $(1.240  \pm 0.037 \pm 0.095)~{\rm GeV}$~\cite{BuchmuellerFlaecher} \\
\hline
\end{tabular}
\caption[Observables]
{Input values and uncertainties used in the CKM fit.}
\label{tab:inputs}
\end{table}

\subsection{Inputs possibly affected by New Physics in mixing}  
The observable $\epsilon_{K}$ has shifted by about $2.3~\%$ (a $3.7~\sigma$ effect) 
between the 2004 and 2006 edition of the PDG from 
$(2.284\pm0.014) \times 10^{-3}$~\cite{PDG2004} down to 
$(2.232\pm0.007) \times 10^{-3}$~\cite{PDG2006}. This shift is mainly caused by 
improved measurements of the branching fraction $BF(K_{L} \to \pi^{+}\pi^{-})$
performed by KTeV, KLOE and NA48 leading to a reduction of $5.5~\%$ of the 
branching fraction values. The translation of $\epsilon_{K}$ into a constraint 
on $\rhobar$ and $\etabar$ suffers from sizeable uncertainties in the decay 
constant ${\hat{B}}_{K}$ (see Table~\ref{tab:inputs}, Ref.~\cite{Tantalo}) and, 
though of less importance, from uncertainties in the QCD corrections coming from 
$\eta_{cc}$~\cite{Nierste1,Nierste2} and from the charm quark mass $m_{c}(m_{c})$ 
in the $\overline{\rm MS}$ scheme obtained from fits to energy and mass moments in 
$B \to X_{c}\ell\nu, X_{s}\gamma$ decays~\cite{BuchmuellerFlaecher}. As discussed 
in Ref.~\cite{BuchmuellerFlaecher} an additional uncertainty of order $50$ MeV 
should be assigned to this value of $m_{c}(m_{c})$ which I add linearly to the 
quoted uncertainty of $45$ MeV from the Heavy Quark Expansion.
Other uncertainties of less importance are coming from $m_{t}$~\cite{Nierste3}, 
and the perturbative QCD corrections $\eta_{tt}$~\cite{Nierste1,Nierste2} and 
$\eta_{ct}$~\cite{Nierste1,Nierste2}.

Within the SM the measurement of the $S$ coefficient in the time-dependent $CP$ 
asymmetry $A_{CP}=S \sin{(\Delta m_{d} \cdot t)} + C \cos{(\Delta m_{d} \cdot t)}$ 
in decays of neutral $B_{d}$ mesons to final states $(c\bar{c}) K^{0}$ provides to 
a very good approximation a measurement of the parameter $\sin{2\beta}$. 
The current uncertainty of $0.025$ is still statistics dominated~\cite{HFAG06}. 
The difference between the measured $S$ coefficient and $\sin{2\beta}$ has been 
estimated in Ref.~\cite{BoosMannelReuter} to be below the $10^{-3}$ level. 
Less stringent constraints on this difference are quoted in 
Refs.~\cite{GrossmanKaganLigeti,LiMishima,Ciuchini1}. 
When interpreting the measured $S$ coefficient as $\sin{(2\beta+2\theta_{d})}$
the $SM4FFC$ hypothesis does not rigourously apply. However, as pointed out in 
Ref.~\cite{AtwoodHiller} the gluonic penguin is $OZI$ suppressed and the 
$Z$-penguin is estimated to be small so that NP in decay is assumed to be 
negligible with respect to the leading tree amplitude. The effect from possible 
NP in $K-\bar{K}$ mixing on $\sin{(2\beta+2\theta_{d})}$ can be neglected as 
well due to the small value of $\epsilon_{K}$.
The measurement on $\sin{(2\beta+2\theta_{d})}$ results in four solutions 
on $\beta+\theta_{d}$. Two out of four solutions can be excluded by measuring 
the sign of $\cos{(2\beta+2\theta_{d})}$. 
For a recent review of \babar\ and Belle measurements see Ref.~\cite{lackerbeauty06}. 
The current experimental results from \babar\ and Belle disfavour negative 
$\cos{(2\beta+2\theta_{d})}$ values but it is considered to be difficult 
by the Heavy Flavour Averaging Group to average the different measurements, 
respectively, to determine a reliable confidence level as a function of 
$\cos{(2\beta+2\theta_{d})}$~\cite{HFAG06}. Hence, as a simplification, it 
is assumed here that $\cos{(2\beta+2\theta_{d})}>0$.

The constraint on $\alpha$ is obtained from time-dependent and 
time-independent measurements in the decays $B \to \pi\pi$, 
$B \to \rho\rho$, and $B \to \rho\pi$.
The time-dependent $CP$ asymmetries measured in $B \to \pi\pi$ 
and also in $B \to \rho\rho$ provide information on the effective 
parameter $\sin{(2\alpha_{eff})}$.
It is possible to translate this measurement into a constraint on
$\alpha$ exploiting isospin symmetry which allows to determine the 
difference $\alpha-\alpha_{eff}$ from data~\cite{GronauLondon}.
Under the assumption of exact isospin symmetry the amplitudes 
$A^{+-} \equiv A(B^{0} \to \pi^{+}\pi^{-})$, 
$A^{00} \equiv A(B^{0} \to \pi^{0}\pi^{0})$, 
and $A^{+0} \equiv A(B^{+} \to \pi^{+}\pi^{0})$ satisfy a triangular 
relationship: $\sqrt{2}A^{+0} - \sqrt{2}A^{00} = A^{+-}$. 
A corresponding relationship holds for the $CP$ conjugated decays: 
$\sqrt{2}{\bar{A}}^{+0} - \sqrt{2}{\bar{A}}^{00} = {\bar{A}}^{+-}$.

The extraction of $\alpha$ from the isospin analysis is independent 
of any possible NP contributions in the $\Delta I=1/2$ decay amplitude 
except for the singular point $\alpha=0$. If there are no NP contributions 
in the $\Delta I=3/2$ decay amplitude the extraction provides 
$\alpha=\pi-\gamma-\beta-\theta_d$.
As a consequence, $\alpha$ is equivalent to $\gamma$ if $\beta+\theta_d$
is measured e.g. from $B \to J/\psi K_{S}$ (with ambiguities).
\begin{figure}[h]
\centering
\includegraphics[width=80mm]{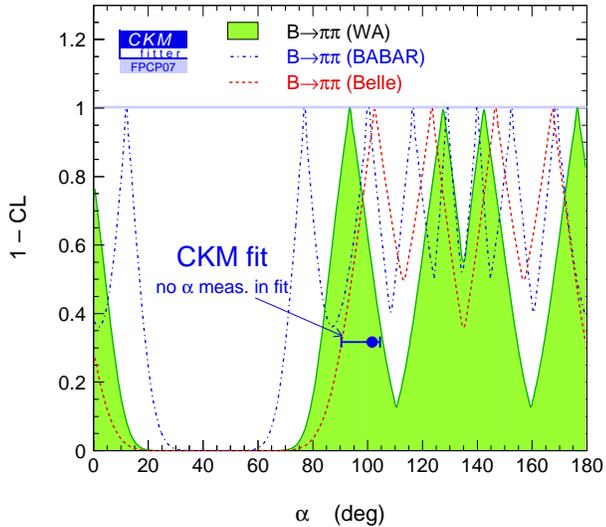}
\caption[alpha from B to pipi]{
Constraints on $\alpha$ from the isospin analysis using $B \to \pi\pi$ data.}
\label{fig:isospinpipi}
\end{figure}

\begin{figure}[h]
\centering
\includegraphics[width=80mm]{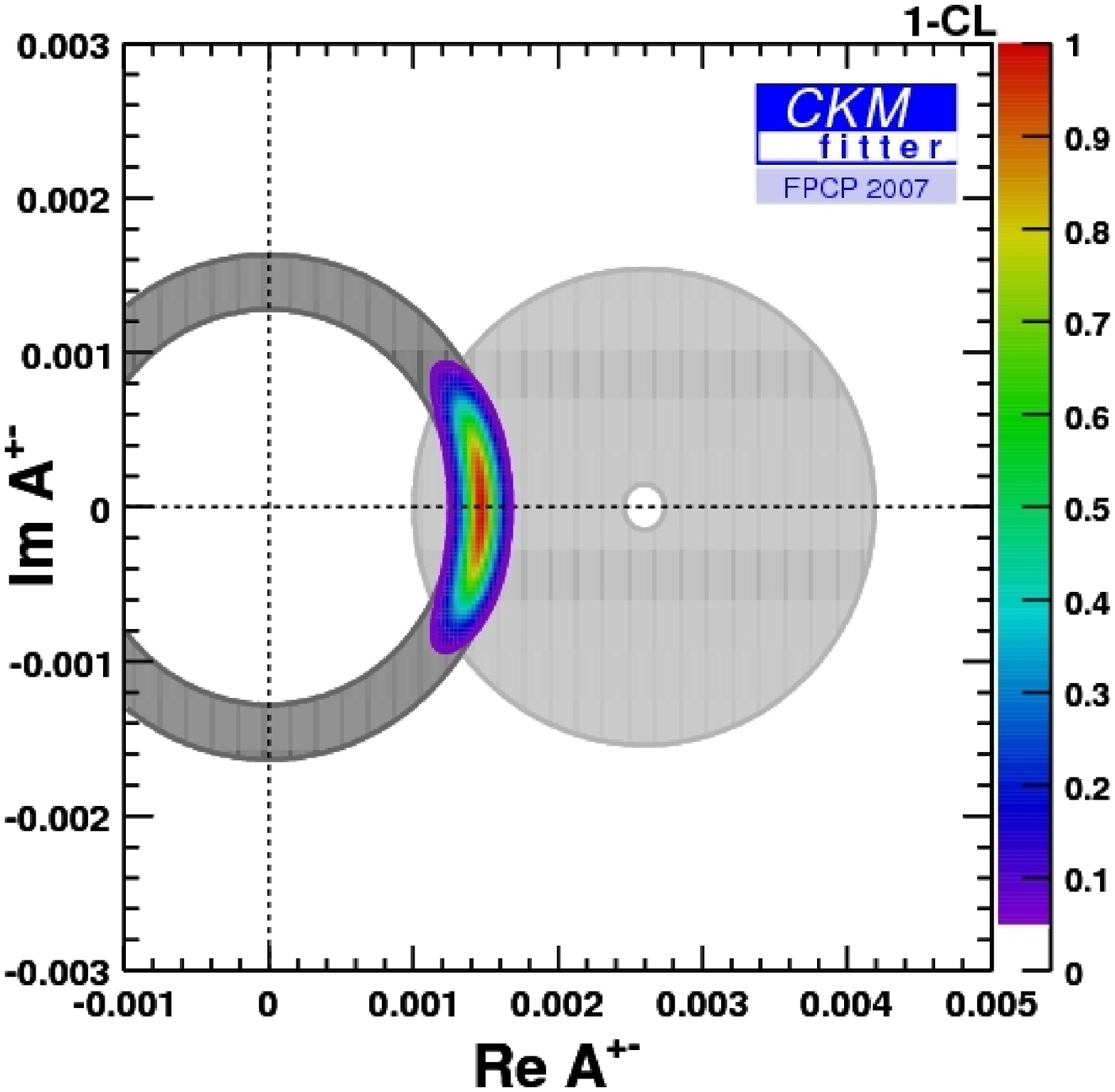}
\includegraphics[width=80mm]{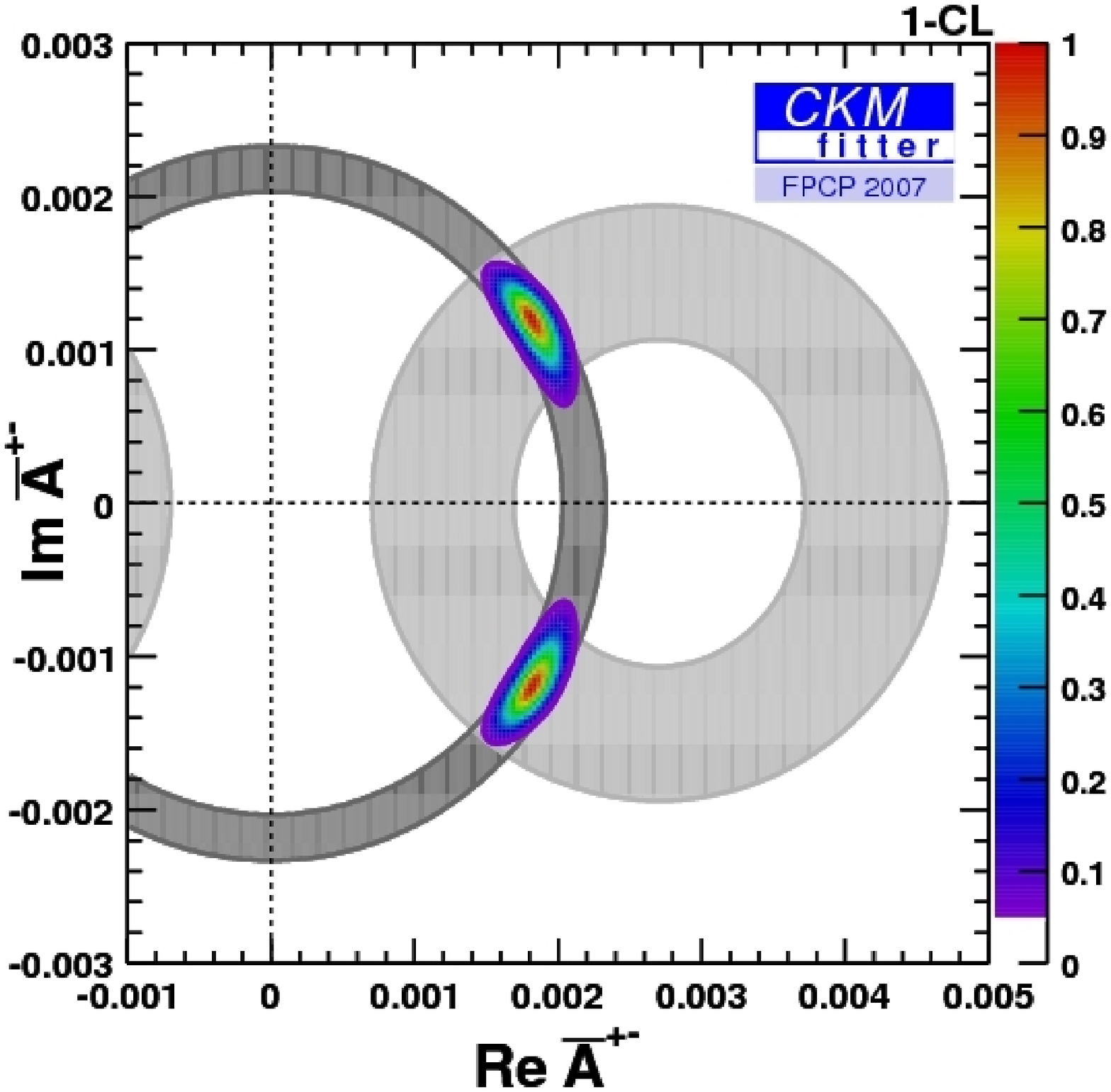}
\caption[Isospin trianglesfor B to pipi]{
The two isospin triangles for the $B \to \pi\pi$ system for the
$B$-meson, respectively, the $\bar{B}$-meson system using the 
world averages~\cite{HFAG06} for branching fractions and direct
$CP$ asymmetries.}
\label{fig:isospintriangles}
\end{figure}

\begin{figure}[h]
\centering
\includegraphics[width=80mm]{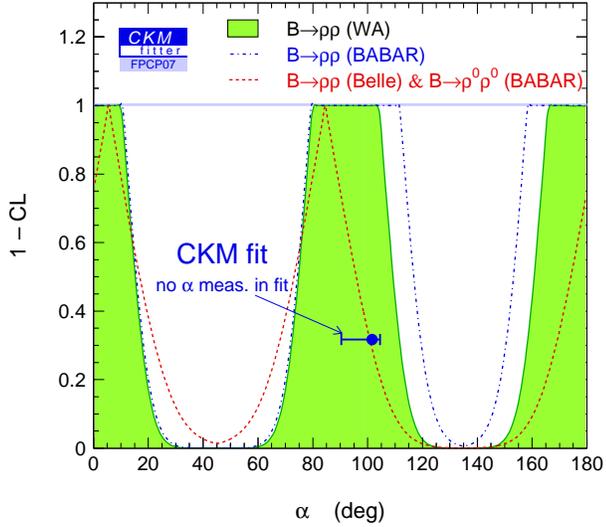}
\caption[alpha from B to rhorho]{
Constraints on $\alpha$ from the isospin analysis using $B \to \rho\rho$ data.}
\label{fig:isospinrhorho}
\end{figure}

The structure seen in the CL as a function of $\alpha$ observed in in the 
isospin analysis (Figs.~\ref{fig:isospinpipi} and~\ref{fig:isospinrhorho}) 
can be easily understood. For $B \to \pi\pi$ the eight solutions from the 
isospin analysis are clearly visible when using only the results for the 
$CP$ asymmetries from \babar\ while for Belle and the world 
averages~\cite{HFAG06} only four solutions are observed. This is due to the 
fact that in the latter case one of the two isospin triangles, the $B$-meson 
triangle, barely closes as illustrated in Fig.~\ref{fig:isospintriangles}.
For $B \to \rho\rho$ there is only evidence so far for the decay 
$B^{0}/{\bar{B}}^{0} \to \rho^{0}\rho^{0}$ from a \babar\ measurement 
but no $CP$ asymmetry had been measured up to this point~\footnote{At
the Lepton-Photon conference 2007, \babar\ has presented for the first 
time a time-dependent $CP$ asmmetry measurement for
$B^{0}/{\bar{B}}^{0} \to \rho^{0}\rho^{0}$ though with still large 
uncertainties on the $S$ and $C$ coefficient.}. 
As a consequence, only a limit on $\alpha-\alpha_{eff}$ can be extracted 
which explains the constant regions in the CL curve. 

Concerning statistical issues in the isospin analysis there has been 
a recent debate in the literature. 
In Ref.~\cite{isospinprior} it has been shown that the result of the 
isospin analysis in $B \to \pi\pi$ and $B \to \rho\rho$ shows a clear 
prior dependence when being performed in the framework of Bayesian 
statistics. 
In a reply, the UTfit collaboration~\cite{replytoisospinprior} argued 
that the result of the isospin analysis for $B \to \pi\pi$ when 
considering $95~\%$ probability intervals shows only a weak prior 
dependence. Moreover, it was advertized that additional information 
on top of the isospin analysis should be used, namely, that SM QCD 
amplitudes can not exceed a certain size in order to avoid $SU(3)$ 
flavour symmetry breaking effects of more than $100~\%$. With this 
additional input the constraint on $\alpha$ are getting stronger and 
the prior dependence becomes weaker.

While there is indeed no large difference between credibility intervals 
with a probability content of $95~\%$ in the $B \to \pi\pi$ analysis 
for the specific data set considered by the UTfit collaboration the 
result of a Bayesian analysis is the full a-posteriori probability 
density function (PDF) often used as input in subsequent analyses 
and not just intervals of a specific probability content as pointed 
out in Ref.~\cite{replytoreplytoisospinprior}. 
In addition, significant differences are observed for e.g. $95~\%$ 
probability intervals when considering the $B \to \rho\rho$ isospin 
analysis. While there is no objection to use additional theory input 
on top of the isospin analysis which improves the constraints on $\alpha$ 
as discussed e.g. in detail in Ref.~\cite{CKMfitter} this kind of 
analysis is not equivalent to the isospin analysis and does not preserve 
the exact degeneracy as expected from the remaining symmetries of the 
problem~\cite{replytoreplytoisospinprior}.

\begin{figure}[h]
\centering
\includegraphics[width=80mm]{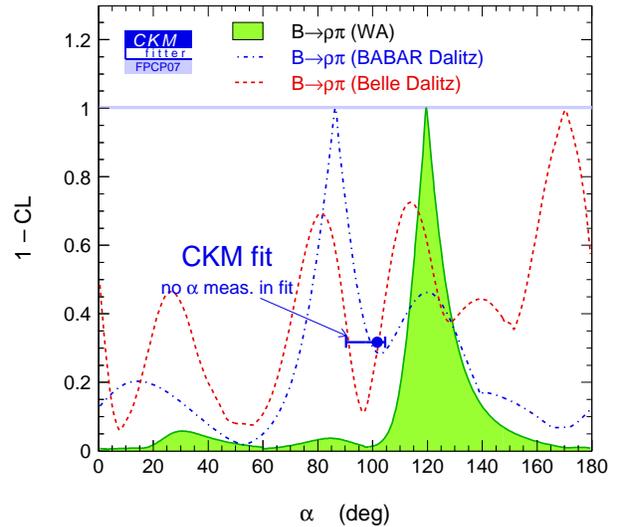}
\caption[alpha from B to rhorho]{
Constraints on $\alpha$ from the $B \to \rho\pi$ Dalitz plot analyses.}
\label{fig:alphaDalitzPlot}
\end{figure}

\begin{figure}[h]
\centering
\includegraphics[width=80mm]{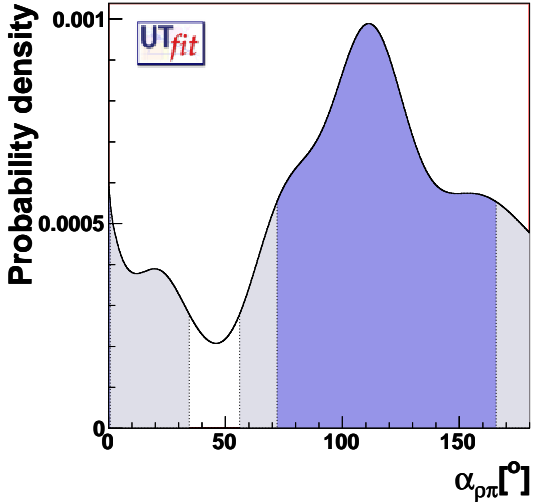}
\caption[alpha from B to rhorho in the UTfit analysis]{
Constraints on $\alpha$ from the $B \to \rho\pi$ Dalitz plot analyses
as performed by the UTfit collaboration in the Summer 2007 analysis.}
\label{fig:alphaDalitzPlotUTfit}
\end{figure}

The UT angle $\alpha$ can also be determined from a $B \to \rho\pi$
Dalitz plot analysis which has the principal advantage that no ambiguities 
are present although, with small statistics, it is possible that mirror 
solutions may occur. 
The \babar\ collaboration has performed such an analysis in 
Ref.~\cite{BabarRhoPiNew} based on a sample of $375 \times 10^{-6}$
$B\bar{B}$ events. A similar analysis but also taking into account
twobody final state $B \to \rho\pi$ branching fractions has been
shown by Belle~\cite{BelleRhoPi} based on a sample of $449 \times 10^{-6}$ 
$B\bar{B}$ events. The constraint on $\alpha$ for the Dalitz plot 
results from \babar\ and Belle as well as for their combination are shown 
in Fig.~\ref{fig:alphaDalitzPlot}. The combined constraint shows the 
particular feature that the occuring mirror solutions have relatively 
small CL values compared to the preferred solution around $120^{\circ}$.
The preferred solution itself deviates from the SM prediction for 
$\alpha$ by about $2\sigma$.
The fact that the combination differs significantly from a naive average
of both experimental CL curves is due to the fact that all the Dalitz 
plot observables, the $U$ and $I$ coefficients, are averaged (and not 
just $\alpha$) including their correlations which is crucial for a 
correct average.
The combined analysis of the UTfit collaboration at the time of the FPCP07
conference, also taking into account the correlation between the $U$ and $I$ 
coefficients, finds a significantly different constraint. One part of the 
discrepancy might stem from a slightly different input data set. The inputs
used by the UTfit collaboration for their Winter 2007 analysis stemmed from 
analyses presented at the ICHEP06 conference: The \babar\ analysis used was 
based on a slightly smaller data set 
($347 \times 10^{6}$ $B \bar{B}$ pairs)~\cite{BabarRhoPiOld} while for the 
Belle analysis the ICHEP06 result based on the same statistics, however, 
with slightly different systematic uncertainties was taken~\cite{BelleRhoPiOld}.
The Summer 2007 UTfit analysis relies on the same inputs as the analysis
presented here and, in contrast to the Winter 2007 analysis, only one preferred 
region around $110^{\circ}$ is observed 
(see Fig.~\ref{fig:alphaDalitzPlotUTfit}) demonstrating the sensitivity of 
the $B \to\rho \pi$ analysis to small changes in the inputs. Nonetheless, the 
global constraint from the UTfit group shows still significant differences 
compared to the Frequentist analysis of the CKMfitter group.

\begin{figure}[h]
\centering
\includegraphics[width=80mm]{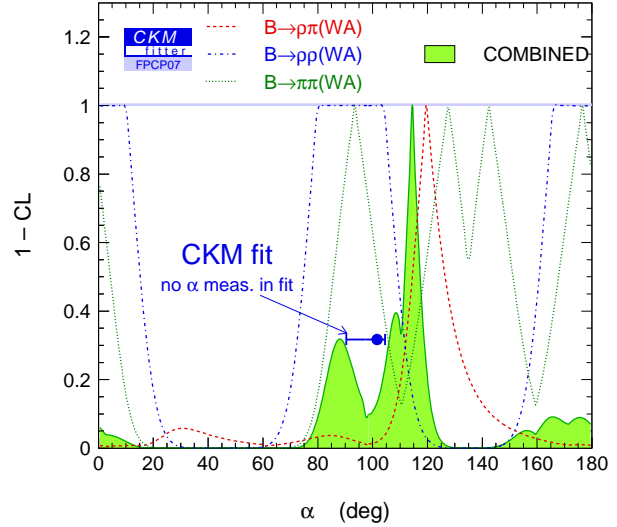}
\caption[alpha from B to rhorho]{
The combined constraint on $\alpha$ from $B \to \pi\pi$, $B \to \rho\rho$
and $B \to \rho\pi$ .}
\label{fig:alphaCombined}
\end{figure}

The combined constraint on $\alpha$ from $B \to \pi\pi$, $B \to \rho\rho$
and $B \to \rho\pi$ is shown in Fig.~\ref{fig:alphaCombined}. Two preferred 
regions are visible around $90^{\circ}$ and $115^{\circ}$ due to the fact 
that the preferred solution from $B \to \rho\pi$ is disfavoured by the 
$B \to \pi\pi$ and $B \to \rho\rho$ constraints while the preferred combined 
$B \to \pi\pi$ and $B \to \rho\rho$ region around $90^{\circ}$ coincides with 
one of the disfavoured $B \to \rho\pi$ solutions. The SM prediction for 
$\alpha$ lies just in between these two regions.
\\

The oscillation frequency in the $B_{d}$ sector $\Delta m_d$ is measured 
with $O(1~\%)$ precision mainly due to the $B$-factory data~\cite{PDG07}. 
Since 2006 $\Delta m_s$ is also known with good precision thanks to the 
observation~\cite{DmsCDF1} of and improved measurement~\cite{DmsCDF2} of 
$B_{s}$ oscillations by CDF.
The translation of the measured value for $\Delta m_q$ into constraints 
on CKM parameters ${|V_{tq} V_{tb}^{*}|}^{2}$ ($q=d,s$) suffer from 
significant uncertainties on $f_{B_{d}} \sqrt{B_d}$, respectively, 
$f_{B_{s}} \sqrt{B_s}$. The input values for these hadronic parameters 
can be calculated in LQCD. For $f_{B_{s}}$, $B_s$, and the ratio 
$f_{B_{s}}/f_{B_{d}}$, the central values and uncertainties are 
used as quoted in a recent review by Tantalo where the first 
uncertainty reflects a statistical error and the second the range 
of various LQCD results~\cite{Tantalo}. 
For the ratio $B_{s}/B_{d}$ Ref.~\cite{Becirevic} is used since 
no corresponding value and uncertainty has been provided in 
Ref.~\cite{Tantalo}.
The value and uncertainty for the perturbative QCD correction 
$\eta_{B}$ is taken from Ref.~\cite{BuchallaBurasLautenbacher}.\\

$CP$ violation in $B_{d}$ mixing ($|q/p| \ne 1$) can be measured 
using the untagged dilepton rate asymmetry~\footnote{With a tagged 
time-dependent decay asymmetry one measures the ratio 
$\frac{1-|q/p|^4}{1+|q/p|^4}$.}
\begin{equation}
A_{SL}^{d}=\frac{N_{\ell^{+}\ell^{+}}-N_{\ell^{-}\ell^{-}} }{N_{\ell^{+}\ell^{+}}+N_{\ell^{-}\ell^{-}}}
          = 2(1-|q/p|).
\end{equation}
Theoretical calculations for $A_{SL}^{d}=\Im{\frac{\Gamma_{12}}{M_{12}}}$ 
at Next-To-Leading Order (NLO) are available~\cite{Ciuchini2,Beneke1,LenzNierste} 
with the most recent one giving a SM prediction of 
$A_{SL}^{d}=(-4.8^{+1.0}_{-1.2}) \times 10^{-4}$~\cite{LenzNierste}. 
As a consequence, a measurement of $A_{SL}^{d}$ is not useful for 
a precise determination of CKM parameters but very helpful to constrain 
possible NP contributions to neutral $B$-meson mixing. A weighted 
average of different results from \babar\, Belle and CLEO results 
in $-0.0043\pm0.0046$~\cite{ASLdBfactories} corresponding to a 
$-0.8~\sigma$ deviation from the SM prediction.
The SM prediction for $A_{SL}^{s}$ in the $B_{s}$ sector is at least 
one order of magnitude smaller than the one for $A_{SL}^{d}$~\cite{LenzNierste}. 
Compared to the SM prediction the first direct measurement from D0 has a quite 
large uncertainty: $A_{SL}^{s}=0.0245\pm0.0196$~\cite{ASLsD0}. D0 has also measured 
an inclusive dimuon asymmetry of $A_{SL}=-0.0028\pm0.0013\pm0.0008$~\cite{ASLD0} 
which is a mixture between the asymmetries $A_{SL}^{d}$ and $A_{SL}^{s}$: 
$A_{SL}=(0.582\pm0.030) A_{SL}^{d}+(0.418\pm0.047) A_{SL}^{s}$~\cite{LenzNierste}.
The measured value corresponds to a $-1.6~\sigma$ deviation from the SM prediction.

A measurement of the lifetime difference in the $B_{s}$ sector provides 
information on the mixing phase $2\phi_{s} = -2 \beta_{s} + 2\theta_s$ 
where the SM prediction
$\beta_{s} \approx \lambda^{2} \eta =(0.945^{+0.201}_{-0.069})^{\circ}~({95~\%~CL})$
is very close to zero with a small uncertainty. In a recent analysis, the 
D0 experiment has measured the untagged time-dependent decay rates for 
$B_{s} \to \psi \phi$ with an angular analysis which allows to disentangle 
the $CP$-even and $CP$-odd final states in this particular vector-vector 
final state~\cite{DeltaGammasCP'}. The result of the analysis is 
$\Delta \Gamma_{s}=(0.17 \pm 0.08(stat)\pm0.02(sys))~{\rm ps^{-1}}$ and 
$\phi_{s}=-0.79\pm 0.56(stat)^{+0.01}_{-0.14}(sys)$\footnote{The result 
contains a four-fold ambiguity which can be reduced to a two-fold ambiguity 
by fixing the strong phase between the $CP$-even and $CP$-odd amplitude as 
it has been done in Ref.~\cite{LenzNierste}. 
Out of the two remaining solutions the one being in good agreement with the 
SM is retained for the further analysis. A different strategy is pursued in 
Ref.\cite{UTfitDeltaF2}.}.
The analysis measures effectively the product 
$\Delta \Gamma^{CP'}_{s} \equiv \Delta \Gamma_{s} \cdot \cos{\phi_s}=\Delta \Gamma_{s}^{SM} \cdot \cos^{2}{\phi_s}$. Taking into account the correlation between the measured $\Delta \Gamma_{s}$
and $\phi_{s}$ results one obtains 
$\Delta \Gamma^{CP'}_{s} = (0.12 \pm 0.08)~{\rm ps^{-1}}$ which is used in 
the NP fit as presented in Sec.~\ref{sec:NPfit}.

\section{The Standard Model CKM fit}  
\label{sec:SMfit}
\begin{figure}[h]
\centering
\includegraphics[width=80mm]{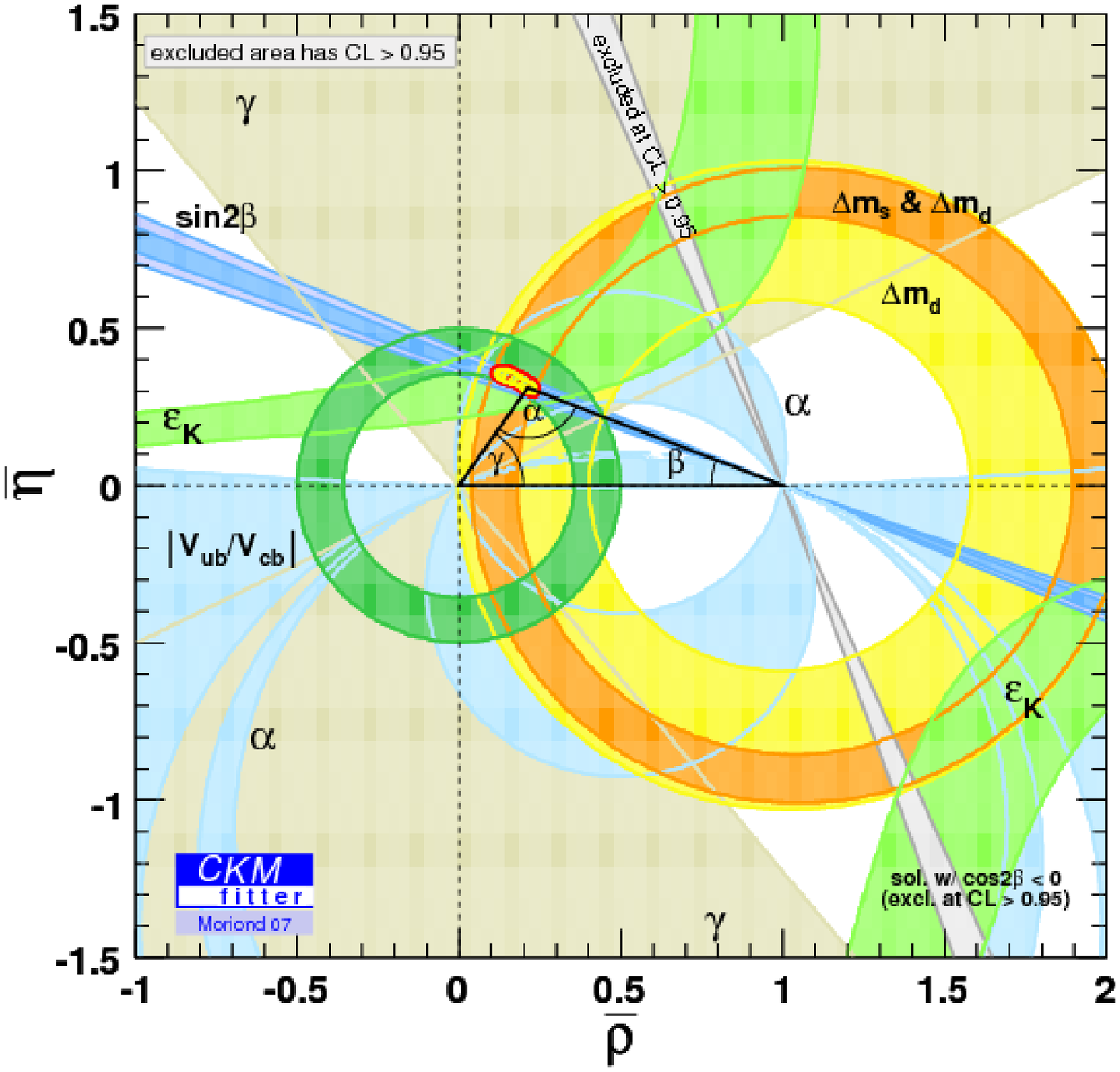}
\caption[rhobar-etabar from SM CKM-fit]{
The constraints on $\rhobar$ and $\etabar$ from the individual constraints
and by combining all individual constraints. Coloured regions indicate
CL's of at least $95~\%$. The allowed region at $95~\%$ CL is shown in 
yellow and inscribed by the red contour. The combined fit also includes 
the input from the branching fraction measurements of $B \to \tau \nu_{\tau}$
as described in Sec.~\ref{sec:BtoTauNu}. This individual constraint is not 
shown in order to guarantee readibility of the plot.
}
\label{fig:rhobaretabar}
\end{figure}

In this section the results of the Standard Model CKM fit are summarized as 
obtained from all inputs quoted in Sec.~\ref{sec:inputs} except the ones for 
$A_{SL}^{d}$, $A_{SL}^{s}$, $A_{SL}$, and from $\Delta \Gamma^{CP'}_{s}$.
The numerical values of the fit results are given in Table~\ref{tab:SMresults}. 
The constraints on $\rhobar$ and $\etabar$ are visualized in Fig.~\ref{fig:rhobaretabar}.
\begin{table}[]
\renewcommand{\arraystretch}{1.3}
\centering
\begin{tabular}{|c|c|}\hline
Parameter                         & Value and Uncertainties ($95~\%$ CL)    \\
\hline
$\lambda_{fit}$                   & $0.2258^{+0.0016}_{-0.0017}$            \\
$A_{fit}$                         & $0.817^{+0.030}_{-0.028}$               \\
$\rhobar_{fit}$                   & $[0.108,0.243]$                         \\
$\etabar_{fit}$                   & $[0.288,0.375]$                         \\
$J_{fit}$                         & $(2.74^{+0.63}_{-0.22}) \times 10^{-5}$ \\
$|V_{ud}^{fit}|$                  & $0.97419\pm0.0037$                      \\
$|V_{us}^{fit}|$                  & $0.2257\pm0.0016$                       \\
$|V_{us}^{pred}|$                 & $0.2275\pm0.0011$                       \\
$|V_{cb}^{fit}|$                  & $(41.7\pm1.3)~10^{-3}$                  \\
$|V_{ub}^{fit}|$                  & $(3.62^{+0.25}_{-0.16}) \times 10^{-3}$ \\
$|V_{ub}^{pred}|$                 & $(3.54^{+0.18}_{-0.16}) \times 10^{-3}$ \\
$|V_{cd}^{fit}|$                  & $0.2255\pm0.0016$                       \\
$|V_{cs}^{fit}|$                  & $0.97334\pm0.00037$                     \\
$|V_{td}^{fit}|$                  & $(8.73^{+0.43}_{-1.14}) \times 10^{-3}$ \\
$|V_{ts}^{fit}|$                  & $(40.9\pm1.3)~10^{-3}$                  \\
$|V_{tb}^{fit}|$                  & $0.999124^{+0.000053}_{-0.000055}$      \\
$\beta_{fit}$                     & $(21.5^{+2.1}_{-1.3})^{\circ}$          \\
$\beta_{pred}$                    & $(26.8^{+2.9}_{-6.2})^{\circ}$          \\
$\alpha_{fit}$                    & $[84.8^{\circ},108.5^{\circ}]$          \\
$\alpha_{pred}$                   & $[85.4^{\circ},107.1^{\circ}]$          \\
$\gamma_{fit}$                    & $[50.7^{\circ},73.1^{\circ}]$           \\
$\gamma_{pred}$                   & $[50.5^{\circ},72.9^{\circ}]$           \\
$\Delta m_{d}^{pred}$             & $(0.42^{+0.33}_{-0.12})~{\rm ps^{-1}}$  \\
$\Delta m_{s}^{pred}$             & $(23.4^{+6.4}_{-8.2})~{\rm ps^{-1}}$    \\
$\beta_{s}^{pred}$                & $(0.9455^{+0.201}_{-0.069})^{\circ}$    \\
$\epsilon_{K}^{pred}$             & $(2.05^{+1.40}_{-0.71}) \times 10^{-3}$ \\
\hline
\end{tabular}
\caption[SM model fit: Numercial results]
{Selected numerical results from the CKM fit within the framework of the SM
 using the inputs as described in the text.}
\label{tab:SMresults}
\end{table}
As a comparison, the Winter 2007 analysis of the UTfit group obtains
for $\rhobar$ and $\etabar$ the following $95~\%$ credibility intervals:
$[0.107,0.222]$, respectively, $[0.307,0.373]$.

\section{Constraints from $B^{+} \to \tau^{+} \nu_{\tau}$}  
\label{sec:BtoTauNu}
The decay $B^{+} \to \tau^{+} \nu_{\tau}$ is interesting for two reasons.
First, if measured with good precision it provides a stringent constraint on 
the product ${|V_{ub}|} \cdot f_{B}$ where $f_{B}$ is the decay constant of 
the charged $B$ meson. Hence, combining  $BF(B^{+} \to \tau^{+} \nu_{\tau})$
with the measurement of $\Delta m_{d}$ allows to remove the dependency on
$f_{B}$ (assuming isospin symmetry $f_{B}=f_{B_{d}}$) when translating these 
measurements into constraints on $\rhobar$ and $\etabar$.

Second, although the decay is mediated at leading order by a tree amplitude
this process is sensitive to NP through a charged Higgs boson exchange.
In a Two-Higgs-Doublet model of type II the prediction of the branching 
fraction~\cite{Hou} is given by
\begin{eqnarray}
BF(B^{+} \to \tau^{+} \nu_{\tau})=  \nonumber \\
BF(B^{+} \to \tau^{+} \nu_{\tau})_{SM} \left( 1-\tan^{2}{\beta}\frac{m_{B}^{2}}{m_{H^{+}}^{2}} \right)
\end{eqnarray}
with the charged Higgs mass $m_{H^{+}}$ and the ratio of the two Higgs 
vacuum expectation values $\tan{\beta}$, and the SM prediction
\begin{eqnarray}
BF(B^{+} \to \tau^{+} \nu_{\tau})_{SM} =   \nonumber \\
\frac{G_{F}^{2}m_{B}}{8\pi}m_{\tau}^{2}\left(1-\frac{m_{\tau}^{2}}{m_{B}^{2}} \right)^{2} f_{B}^{2}{|V_{ub}|}^{2} \tau_{B}= \nonumber \\
(0.96^{+0.38}_{-0.20}) \times 10^{-4}~(95~\%CL)
\end{eqnarray}
using $|V_{ub}^{fit}|=(3.62^{+0.25}_{-0.16}) \times 10^{-3}$ from the 
CKM fit result as shown in Table~\ref{tab:SMresults} and 
$f_{B}=f_{B_{d}}=(226\pm15\pm26)~{\rm MeV}$ following from the inputs
provided in Sec.~\ref{tab:inputs}.

\begin{figure}[h]
\centering
\includegraphics[width=80mm]{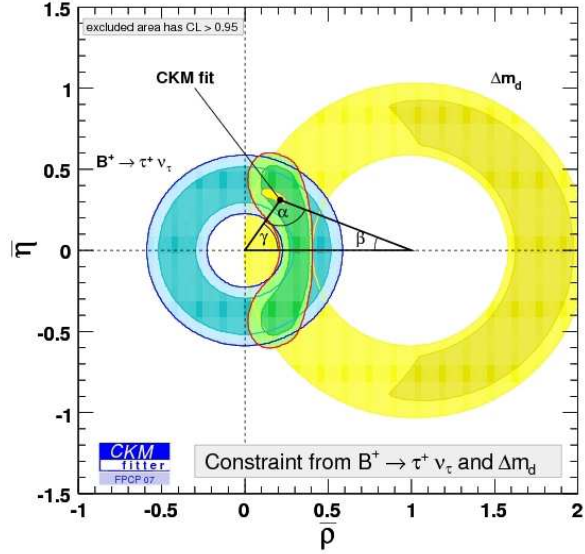}
\caption[rhobar-etabar from BtoTauNu and Dmd]{
Constraints on $\rhobar$ and $\etabar$ from the world averages of 
$BF(B^{+} \to \tau^{+} \nu_{\tau})$ (blue) and $\Delta m_{d}$ (yellow)
and when both are combined (green).}
\label{fig:rhobaretabarfromBtoTauNuandDmd}
\end{figure}

\begin{figure}[h]
\centering
\includegraphics[width=80mm]{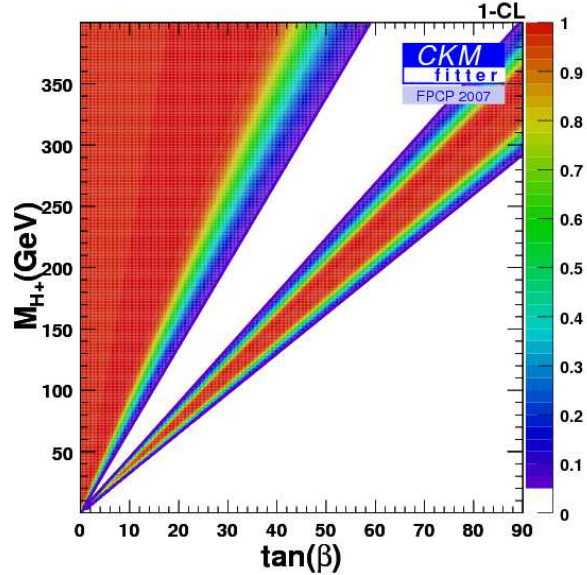}
\caption[Constraints on Charged Higgs from BtoTauNu]{
Constraints from $BF(B^{+} \to \tau^{+} \nu_{\tau})$ on $\tan{\beta}$ 
and $m_{H^{+}}$ in a Two-Higgs-Doublet model of type II using the values 
for $f_{B}=f_{B_{d}}$ and $|V_{ub}|$ as quoted in Sec.~\ref{sec:inputs}.}
\label{fig:ChargedHiggsfromBtoTauNu}
\end{figure}

At this conference the \babar\ collaboration has presented new results for 
the branching fraction $BF(B^{+} \to \tau^{+} \nu_{\tau})$~\cite{Gritsan}. 
The combined result of analyses where the other $B$ meson is tagged by either 
reconstrucing a hadronic decay or a $B \to D^{(*)} \ell \nu$ decay is 
$BF(B^{+}\to\tau^{+}\nu_{\tau})=(1.20^{+0.40+0.29}_{-0.38-0.30}\pm0.22)\times 10^{-4}$
corresponding to a $2.6~\sigma$ access~\cite{Gritsan}.
Belle had already reported a $3.5~\sigma$ evidence with hadronic tags only
and measured
$BF(B^{+}\to\tau^{+}\nu_{\tau})=(1.79^{+0.56+0.46}_{-0.49-0.51})\times 10^{-4}$~\cite{BelleTauNu}.
Fig.~\ref{fig:rhobaretabarfromBtoTauNuandDmd} shows the constraint from 
$BF(B^{+} \to \tau^{+} \nu_{\tau})$ in blue and from $\Delta m_{d}$ in 
yellow. The green region represents the $95~\%$ CL region when both inputs
are combined illustrating the correlation between 
$BF(B^{+} \to \tau^{+} \nu_{\tau})$ and $\Delta m_{d}$ due to the decay 
constant $f_{B}$. 
At present, the precision of $BF(B^{+} \to \tau^{+} \nu_{\tau})$ is not 
sufficient to compete with the combined SM CKM fit.
In Fig.~\ref{fig:ChargedHiggsfromBtoTauNu} are shown the constraints from 
$BF(B^{+} \to \tau^{+} \nu_{\tau})$ on $\tan{\beta}$ and $m_{H^{+}}$ 
in a Two-Higgs-Doublet model of type II using the values for $f_{B}=f_{B_{d}}$ 
and $|V_{ub}|$ as quoted in Sec.~\ref{sec:inputs}.

\section{Constraints on New Physics in $B-\bar{B}$ mixing}  
\label{sec:NPfit}
\begin{figure}[h]
\centering
\includegraphics[width=80mm]{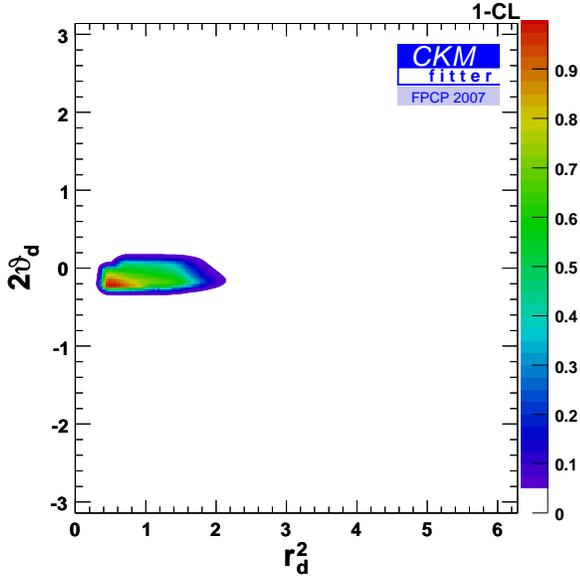}
\caption[NP in Bd mixing: rd2-2thetad]{
Constraints on the NP parameters $r_{d}^2$ and $2 \theta_{d}$ from
a combined fit allowing for NP in $B_d$ and $B_s$ mixing using all 
inputs listed in Table~\ref{sec:inputs}.}
\label{fig:rd2-2thetad-all}
\end{figure}

\begin{figure}[h]
\centering
\includegraphics[width=80mm]{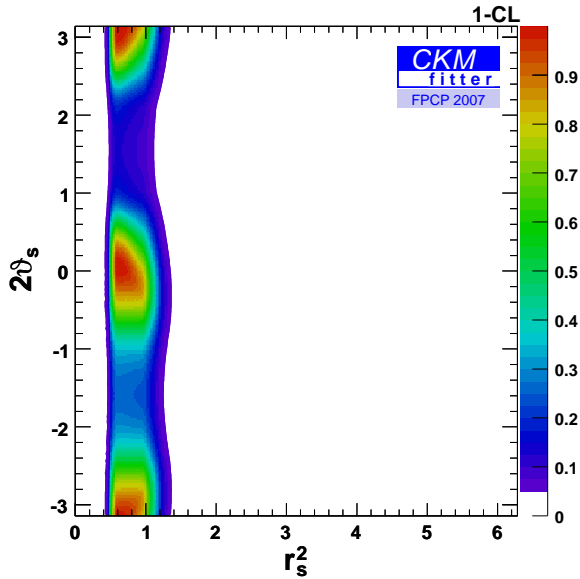}
\caption[NP in Bs mixing: rs2-2thetas]{
Constraints on the NP parameters $r_{s}^2$ and $2 \theta_{s}$ from
a combined fit allowing for NP in $B_d$ and $B_s$ mixing using all 
inputs listed in Table~\ref{sec:inputs}.}
\label{fig:rs2-2thetas-all}
\end{figure}

\begin{figure}[h]
\centering
\includegraphics[width=80mm]{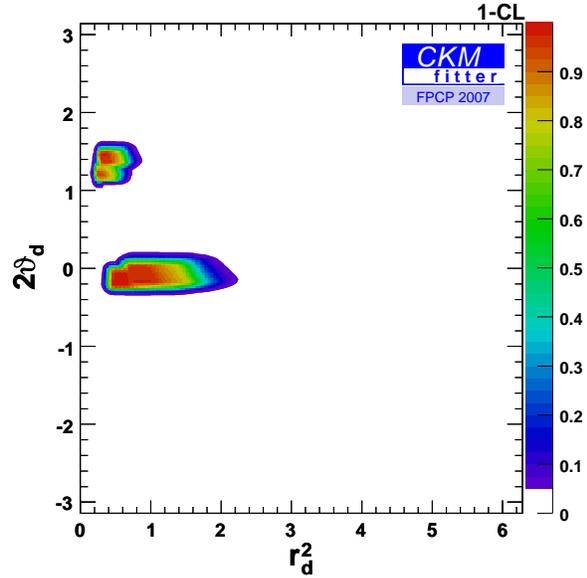}
\caption[NP in Bd mixing: rd2-2thetad wo ASL]{
Constraints on the NP parameters $r_{d}^2$ and $2 \theta_{d}$ from
a combined fit allowing for NP in $B_d$ and $B_s$ mixing using all 
inputs except $A_{SL}^{d}$ and $A_{SL}$.}
\label{fig:rd2-2thetad-woasl}
\end{figure}

When performing a combined fit without using $\epsilon_{K}$ as an 
input and allowing for NP in $B_d$ and $B_s$ mixing one obtains the 
constraints on the NP parameters $r_{q}^2$ and $2 \theta_{q}$ as 
shown in 
Figs.~\ref{fig:rd2-2thetad-all} and ~\ref{fig:rs2-2thetas-all}.
For both neutral $B$-meson systems the SM point showsa decent CL with
the current data although the best fit values in the $B_{d}$ case prefer 
small negative $2 \theta_{d}$ values mainly caused by the inputs from 
$|V_{ub}|$ and $\sin{(2\beta)}$. Also $r_{d}^2$ values smaller than 1 
are preferred due to the slight discrepancy between the $\alpha$ 
constraint and the $\alpha$ SM prediction.
The constraints on $r_{d}^2$ and $2 \theta_{d}$ profit from the inputs 
for $A_{SL}^{d}$ and $A_{SL}$ highlighting the importance of the D0 
measurement and the nice interplay between these inputs from the 
$B$-factories on one hand and the Tevatron on the other hand. 
The preferred large negative values for both, $A_{SL}^{d}$ and $A_{SL}$, 
disfavour large positive $2 \theta_{d}$ values and at the same time 
$r_{d}^2$ values smaller than 1. This is best illustrated in 
Fig.~\ref{fig:rd2-2thetad-woasl} where all inputs but $A_{SL}^{d}$ and 
$A_{SL}$ are used. In this case a second allowed region appears which 
is inconsistent with the SM point. The double-peak structure observed 
is caused by the corresponding structure in the $\alpha$ input.
To pin down a possible significant deviation from $2 \theta_{d}=0$ in the 
future depends significantly on the central value and and the prospects for 
the uncertainty on $\sin{(2\beta)}$ and, more importantly, on $|V_{ub}|$.
To reduce the allowed size for $r_{d}^2$ requires improved precision on 
$\gamma$ and/or $\alpha$, but even more important is an improvement for 
(LQCD) calculations of $f_{B_{d}} \sqrt{B_d}$.

The allowed range in $r_{s}^2$ is significantly smaller than the one for $r_{d}^2$ 
thanks to the more precise value for $f_{B_{s}} \sqrt{B_s}$ compared to
$f_{B_{d}} \sqrt{B_d}$ . Currently, there is no exclusion better than at $90~\%$ CL 
for the NP phase $2 \theta_{s}$. 
The fact that values for $2 \theta_{s}$ around $\pm 90^{\circ}$ are disfavoured 
are due to the fact that the measured value for $\Delta \Gamma_{s}^{CP'}$ is 
larger than the SM prediction while NP ($\propto \cos^{2}{2 \theta_{s}}$) can 
only lower the measured value with respect to the SM prediction. 
With improved measurements of $\Delta \Gamma_{s}^{CP'}$ and/or $\Delta \Gamma_{s}$ 
from Tevatron and eventually from LHCb the SM value $2 \theta_{s}=0$ might be 
excluded. However, the best sensitivity to $2 \theta_{s}$ will only come from 
a tagged time-dependent angular analysis of the decay $B_{s} \to \psi \phi$ on 
a high statistics sample eventually carried out by LHCb and possibly also by 
ATLAS and CMS.

The preferred value in $r_{s}^2$ is smaller than 1 although the difference in 
CL with respect to $r_{s}^2=1$ is small. The current situation with preferred 
values for both, $r_{d}^2$ and $r_{s}^2$, equal but smaller than 1 is consistent 
with a Minimal Flavour Violation scenario.

\begin{figure}[h]
\centering
\includegraphics[width=80mm]{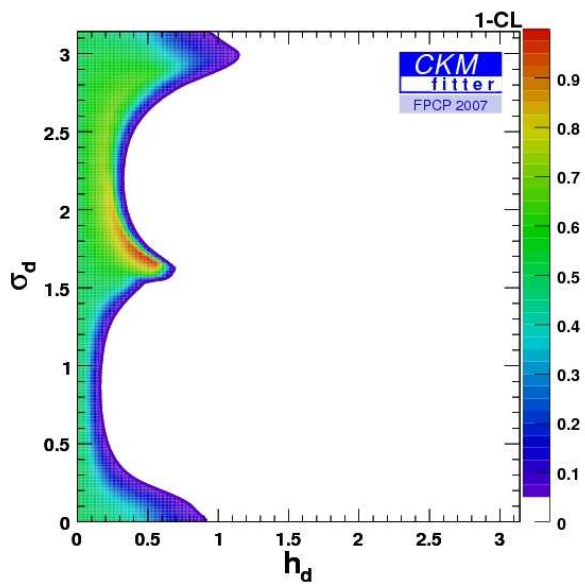}
\caption[NP in Bs mixing: hd-2sigmad]{
Constraints on the NP parameters $h_{d}$ and $2 \sigma_{d}$ from
a combined fit allowing for NP in $B_d$ and $B_s$ mixing using all 
inputs listed in Table~\ref{sec:inputs}.}
\label{fig:hd-2sigmad-all}
\end{figure}

\begin{figure}[h]
\centering
\includegraphics[width=80mm]{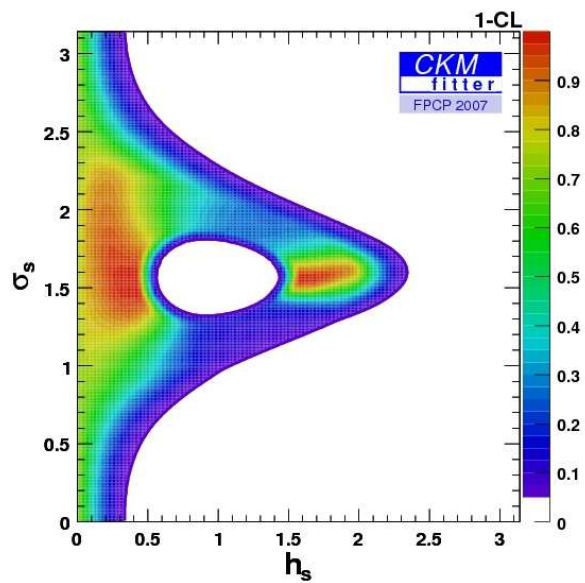}
\caption[NP in Bs mixing: hs-2sigmas]{
Constraints on the NP parameters $h_{s}$ and $2 \sigma_{s}$ from
a combined fit allowing for NP in $B_d$ and $B_s$ mixing using all 
inputs listed in Table~\ref{sec:inputs}.}
\label{fig:hs-2sigmas-all}
\end{figure}

Constraints in  $h_{d}$ and $2 \sigma_{d}$, respectively,  $h_{s}$ and $2 \sigma_{s}$
corresponding to Figs.~\ref{fig:rd2-2thetad-all} and ~\ref{fig:rs2-2thetas-all}.
are shown in Figs.~\ref{fig:hd-2sigmad-all} and ~\ref{fig:hs-2sigmas-all}.
\\

Similar in line, the constraint on $h_{K}$ and $2\sigma_{K}$ from a combined
analysis in the $K-\bar{K}$-, $B_{d}-{\bar{B}}_{d}$- and 
$B_{s}-{\bar{B}}_{s}$-system as presented in Ref.~\cite{Agashe} shows that 
all $2\sigma_{K}$ values are allowed.
Except for $2\sigma_{K}$ around $70^{\circ}$ and $155^{\circ}$ the parameter 
$2\sigma_{K}$ is constrained to be smaller than $\approx 0.55$ at $95~\%$ CL.
According to the conclusions of Ref.~\cite{LigetiPapucciPerez} 
the current constraints on the $K-\bar{K}$-, $B_{d}-{\bar{B}}_{d}$- and 
$B_{s}-{\bar{B}}_{s}$-system are still compatible with a NMFV 
scenario since the sufficiently large $h_{q}$ values ($q=K,d,s$) are still 
allowed~\footnote{For a recent work on the scale of certain NMFV models 
see Ref.~\cite{UTfitDeltaF2}.}.

\begin{acknowledgments}
I like to thank the organizers for the invitation to this very beautiful 
conference. I profited in particular from discussions with P.~Ball, 
D.~Becirevic, M.~Gronau, W.S.~Hou, A.~Lenz, H.-nan Li, Z.~Ligeti, M.~Neubert, 
U.~Nierste, M.~Papucci, and G.~Perez. 
My warmest thanks go to my collaborators of the CKMfitter group for their 
support in the preparation of this talk.
\end{acknowledgments}

\bigskip 

\end{document}